\documentstyle[12pt]{article}
\font\twelve=cmbx10 at 15pt
\font\ten=cmbx10 at 12pt

\begin{document}

\begin{titlepage}

\begin{center}

\renewcommand{\thefootnote}{\fnsymbol{footnote}}

{\ten Centre de Physique Th\'eorique\footnote{
Unit\'e Propre de Recherche 7061} - CNRS - Luminy, Case 907}
{\ten F-13288 Marseille Cedex 9 - France }

\vspace{1 cm}

{\twelve QUANTUM EFFECTS ASSOCIATED WITH AN ELECTRICALLY NEUTRAL MASSLESS
PARTICLE IN A BERTOTTI-ROBINSON UNIVERSE}

\vspace{0.3 cm}

\setcounter{footnote}{0}
\renewcommand{\thefootnote}{\arabic{footnote}}

{\bf V\'{\i}ctor M. VILLALBA\footnote{permanent address: Centro de
F\'{\i}sica, Instituto Venezolano de Investigaciones Cient\'{\i}ficas IVIC,
Apdo. 21827, Caracas 1020-A Venezuela\\ e-mail
address: villalba@cpt.univ-mrs.fr, villalba@dino.conicit.ve} }

\vspace{2 cm}

{\bf Abstract}

\end{center}

In the present article we obtain, by separation of variables, an exact
solution to the Dirac equation with anomalous momentum for an electrically
neutral massless particle in  a Bertotti-Robinson universe.  We discuss the
phenomenon of particle creation in this model.

\vspace{2,5 cm}


\bigskip

\noindent October 1994

\noindent CPT-94/P.3081

\bigskip

\noindent anonymous ftp or gopher: cpt.univ-mrs.fr

\end{titlepage}

In the past twenty years there has been a increasing interest in studying
quantum effects in cosmological backgrounds. Since the publication of the
pioneer works of Parker \cite{parker} and Grib {\it et al }\cite{Grib} , a
great body of articles have been devoted to the problem of studying the
creation of scalar and spin 1/2 particles in different cosmological
scenarios. In this direction the formulation of a quantum field theory on
curved manifolds is required, and consequently we have to solve in the
corresponding curved background the relativistic wave functions associated
with one-particle states.

Despite the great effort dedicated to the search of exact solutions of
relativistic wave equations in curved space-times, the number of exact
solvable examples is relatively scarce. This one, in most of the cases, is
related to the impossibility of achieving the preliminary step of separating
variables in the corresponding wave equations. Therefore we have to restrict
ourselves to those metrics permitting such a separation of variables. That
is the case for some diagonal metric, as well as the type D Petrov spaces.
The presence of conserved quantities as well as a high degree of symmetries
of the metric are desirable if we are willing to find exact solution of the
Klein-Gordon or the Dirac equations. This is the case for homogeneous spaces
like the Friedmann-Robertson-Walker metric.

A good scenario for discussing quantum effects is the Bertotti-Robinson
electromagnetic universe \cite{Bertotti,Robinson,sahni}. This is a
non-asymptotically flat space-time generated by a constant, homogeneous
electric field. The metric associated with this universe can be written as
\begin{equation}
\label{1}ds^2=\frac 1{\sqrt{\lambda }}\left[ d\vartheta ^2+\sin ^2\vartheta
d\varphi ^2+\cos ^2tdR^2-dt^2\right]
\end{equation}
Among the advantages of the metric (\ref{1}) we can mention that it can be
regarded as a reducible Einstein space , also it has zero Weyl tensor and ,
because of its diagonal and very simple form, the Klein Gordon and Dirac
equations are exactly soluble in terms of special functions\cite{sahni}.
Perhaps the Bertotti Robinson universe is the only model available in the
literature where it is possible to solve the wave equations taking into
account that the background field is charged and therefore is present the
electromagnetic minimal coupling via the vector potential $A_\mu $. Here it
is worth mentioning that is not the case for the cosmological models where
the electromagnetic interaction goes in the radiation fluid.

Since the metric (\ref{1}) is conformally flat, creation of electric free
massless particles cannot be expected \cite{Zeldovich}. This situation can
be modified if we consider the existence of a no trivial anomalous moment.
In this case, there is an additional term in the massless Dirac equation
\begin{equation}
\label{2}\left( \gamma ^\alpha (\partial _\alpha -\Gamma _\alpha )+\xi
\gamma ^\alpha \gamma ^\beta F_{\alpha \beta }\right) \Psi =0
\end{equation}
due to the presence of the anomalous electric moment. It is purpose of this
letter to discuss the mechanism of particle creation of electrically neutral
massless spin 1/2 particles in the background field (\ref{1})

The electromagnetic field, source of the gravitational field with the metric
(\ref{1}) can be written as
\begin{equation}
\label{3}A_R=\frac 1{\lambda ^{1/4}\chi ^{1/2}}\sin t,\quad
F_{01}=F_{tR}=\frac 1{\lambda ^{1/4}\chi ^{1/2}}\cos t
\end{equation}
and, the curved Dirac matrices $\gamma ^\mu $ can be expressed, in the
diagonal (rotating) tetrad gauge, in terms of the flat gamma's $\tilde
\gamma ^\mu $ as follows,
\begin{equation}
\label{4}\gamma ^0=\lambda ^{1/4}\tilde \gamma ^0,\ \gamma ^1=\frac{\lambda
^{1/4}}{\cos t}\tilde \gamma ^1,\ \gamma ^2=\lambda ^{1/4}\tilde \gamma ^2,\
\gamma ^3=\frac{\lambda ^{1/4}}{\sin \vartheta }\tilde \gamma ^3
\end{equation}
then, substituting the gamma matrices $\gamma ^\mu $ into the expression for
the spinor connections $\Gamma _\alpha $
\begin{equation}
\label{5}\Gamma _\alpha =\frac 14g_{\mu \lambda }\left[ \left( \frac{%
\partial b_\nu ^\beta }{\partial x^\alpha }\right) a_\beta ^\lambda -\Gamma
_{\nu \alpha }^\lambda \right] s^{\mu \nu }
\end{equation}
where
\begin{equation}
\label{6}s^{\mu \nu }=\frac 12(\gamma ^\mu \gamma ^\nu -\gamma ^\nu \gamma
^\mu )
\end{equation}
and the matrices $b_\alpha ^\beta $ , $a_\beta ^\alpha $ establish the
connection between the curved Dirac matrices $\gamma $ and the constant
Dirac matrices $\tilde \gamma $ as follows
\begin{equation}
\label{7}\gamma _\mu =b_\mu ^\alpha \tilde \gamma _\alpha ,\quad \gamma ^\mu
=a_\beta ^\mu \tilde \gamma ^\beta
\end{equation}
we obtain
\begin{equation}
\label{8}\Gamma _0=0,\ \Gamma _1=-\frac 12\sin t\tilde \gamma ^0\tilde
\gamma ^1,\ \Gamma _2=0,\ \Gamma _3=\frac 12\cos \vartheta \tilde \gamma
^2\tilde \gamma ^3,
\end{equation}
Substituting the connections $\Gamma _\alpha $ into the Dirac equation, and
taking into account (\ref{3}) and (\ref{4}) we arrive at
\begin{equation}
\label{9}\left( \tilde \gamma ^0\partial _t+\frac{\tilde \gamma ^1}{\cos t}%
\partial _R+\tilde \gamma ^2\partial _\vartheta +\frac{\tilde \gamma ^3}{%
\sin \vartheta }\partial _\varphi +\frac{i\xi }{\chi ^{1/2}}\tilde \gamma
^0\tilde \gamma ^1\right) \Phi =0
\end{equation}
where we have introduced the auxiliary spinor $\Phi $%
\begin{equation}
\label{10}\Psi =(\cos t\sin \vartheta )^{-1/2}\Phi
\end{equation}
Applying the algebraic method of separation of variables \cite
{Shishkin1,Shishkin2}, it is straightforward to separate the angular
dependence of $\Phi $ in (\ref{9}) from the radial and time ones. Then,
after introducing the spinor $\Theta =\tilde \gamma ^0\tilde \gamma ^1\Phi $
we have that eq. (\ref{9}) reduces to
\begin{equation}
\label{K}-i\left( \tilde \gamma ^2\partial _\vartheta +\frac{\tilde \gamma ^3%
}{\sin \vartheta }\partial _\varphi \right) \tilde \gamma ^0\tilde \gamma
^1\Theta =k\Theta
\end{equation}
\begin{equation}
\label{11}\left( -\tilde \gamma ^1\partial _t-\frac{\tilde \gamma ^0}{\cos t}%
\partial _R+\frac{i\xi }{\chi ^{1/2}}+ik\right) \Theta =0
\end{equation}
Here, some words about the chirality condition
\begin{equation}
\label{ch}(1-i\gamma _5)\Phi =0
\end{equation}
are in order, it is not difficult to see that neither the momentum operator
in (\ref{K}) nor eq. (\ref{11}) commute with $(1-i\gamma _5)$, then, in
order preserve the condition (\ref{ch}) we shall consider as solution of our
problem the expression $\Phi _c=$ $(1+i\gamma _5)\Phi $ which obviously
satisfies eq. (\ref{ch}).

Noticing that the operator acting on $\Theta $ in eq. (\ref{K}) is the Brill
and Wheeler angular momentum operator \cite{Brill}, we have that the
constant of separation $k$ takes integer values $k=\pm 1,\pm 2...$ Using the
Dirac matrices representation
\begin{equation}
\label{12}\gamma ^0=\left(
\begin{array}{cc}
0 & i\sigma ^2 \\
i\sigma ^2 & 0
\end{array}
\right) ,\ \gamma ^1=\left(
\begin{array}{cc}
0 & \sigma ^1 \\
\sigma ^1 & 0
\end{array}
\right) \gamma ^2=\left(
\begin{array}{cc}
-1 & 0 \\
0 & 1
\end{array}
\right) ,\quad \gamma ^3=\left(
\begin{array}{cc}
0 & \sigma ^3 \\
\sigma ^3 & 0
\end{array}
\right)
\end{equation}

we have eq. (\ref{11}) can be reduced to the form
\begin{equation}
\label{13}\left( \frac d{dt}-\frac{ik_R}{\cos t}\right) \left(
\begin{array}{c}
\Theta _{11} \\
\Theta _{21}
\end{array}
\right) -i\left( \frac \xi {\chi ^{1/2}}+k\right) \left(
\begin{array}{c}
\Theta _{22} \\
\Theta _{12}
\end{array}
\right) =0
\end{equation}
\begin{equation}
\label{14}\left( \frac d{dt}+\frac{ik_R}{\cos t}\right) \left(
\begin{array}{c}
\Theta _{12} \\
\Theta _{22}
\end{array}
\right) -i\left( \frac \xi {\chi ^{1/2}}+k\right) \left(
\begin{array}{c}
\Theta _{21} \\
\Theta _{11}
\end{array}
\right) =0
\end{equation}
where
\begin{equation}
\label{tita}\Theta =\left(
\begin{array}{c}
\Theta _{11} \\
\Theta _{12} \\
\Theta _{21} \\
\Theta _{22}
\end{array}
\right)
\end{equation}
the system of equations takes a more familiar form after making the change
of variables
\begin{equation}
\label{15}t=z+\frac \pi 2,\ \cos t=-\sin z,
\end{equation}
then, we look for solutions of the system (\ref{13}) in the form
\begin{equation}
\label{16}\Theta _{11}=\sin ^{ik_R}(z)\sin (\frac z2)I(z),\quad \Theta
_{22}=\sin ^{ik_R}(z)\cos (\frac z2)J(z)
\end{equation}
where the functions $I$ and $J$ satisfy the system of equations
\begin{equation}
\label{17}(q+1)\frac{dI}{dq}+(ik_R+\frac 12)I=-i\left( \frac \xi {\chi
^{1/2}}+k\right) J
\end{equation}
\begin{equation}
\label{18}(q-1)\frac{dJ}{dq}+(ik_R+\frac 12)J=i\left( \frac \xi {\chi
^{1/2}}+k\right) I
\end{equation}
and the new auxiliary variable $q$ reads
\begin{equation}
\label{19}q=\sin t
\end{equation}
The solution of the system (\ref{17})-(\ref{18}) can be expressed in terms
of hypergeometric functions $F(a,b,\gamma ,z)$ as follows:
\begin{equation}
\label{20}J=sF(a,b,ik_R+\frac 12,(q+1)/2)
\end{equation}
\begin{equation}
\label{21}I=-i\frac{\frac \xi {\chi ^{1/2}}+k}{ik_R+\frac 12}%
sF(a,b,ik_R+\frac 32,(q+1)/2)
\end{equation}
where the parameters $a$ and $b$ read
\begin{equation}
\label{22}a=ik_R+\frac 12+\left( \frac \xi {\chi ^{1/2}}+k\right) ,\quad
b=ik_R+\frac 12-\left( \frac \xi {\chi ^{1/2}}+k\right)
\end{equation}
In order to analyze the phenomenon of particle creation in the background
field (\ref{1}) we proceed to discuss the behavior of the solutions of the
Dirac equation in the hypersurfaces $t=\pm \pi /2,$ or equivalently when $%
q=\pm 1.$ We can establish the asymptotic behavior of the solutions at the
asymptotes by making a comparison with the quasiclassical behavior
associated with the Hamilton-Jacobi equation,
\begin{equation}
\label{23}g^{\alpha \beta }\frac{\partial S}{\partial x^\alpha }\frac{%
\partial S}{\partial x^\beta }=0
\end{equation}
since the metric $g^{\alpha \beta }$ associated with (\ref{1}) does not
depend on the variables $R$ and $\varphi ,$ we have that the function $S$
can be separated as follows,
\begin{equation}
\label{24}S=K_RR+k_\varphi \varphi +S_t(t)+S_\vartheta (\vartheta )
\end{equation}
then, substituting (\ref{24}) into (\ref{23}) we arrive at
\begin{equation}
\label{25}-\left( \frac{dS_t}{dt}\right) ^2+\frac{K_R^2}{\cos ^2t}+\left(
\frac{dS_\vartheta }{d\vartheta }\right) ^2+\frac{k_\varphi ^2}{\sin
^2\vartheta }=0
\end{equation}
whose solution can be written as
\begin{equation}
\label{St}\left( \frac{dS_t}{dt}\right) =\pm \sqrt{\frac{K_R^2}{\cos ^2t}%
+\lambda ^2}
\end{equation}
where $\lambda $ is a constant of separation associated with the angular
dependence of the function S. It is not difficult to see that for $t=\pm \pi
/2$ the expression (\ref{St}) reduces to
\begin{equation}
\label{26}S_t=\pm K_R\log \mid \sec t+\tan t\mid +C,\quad
\end{equation}
then we have that for $q\rightarrow 1$ $(1-q)^{-ik_R}$ and $(1-q)^{+ik_R}$ ,
and for $q\rightarrow -1$ $(1+q)^{ik_R}$ and $(1-q)^{-ik_R}$ correspond to
negative and positive frequency solutions respectively. In order to relate
the two different vacua for $q=\pm 1$, we can make use of the relation for
the hypergeometric functions $F(\alpha ,\beta ,\gamma ,z)$
$$
F(\alpha ,\beta ,\gamma ,z)=\frac{\Gamma (\gamma )\Gamma (\gamma -\alpha
-\beta )}{\Gamma (\gamma -\alpha )\Gamma (\gamma -\beta )}F(\alpha ,\beta
,\alpha +\beta -\gamma +1,1-z)+
$$
\begin{equation}
\label{27}+\frac{\Gamma (\gamma )\Gamma (\alpha +\beta -\gamma )}{\Gamma
(\alpha )\Gamma (\beta )}(1-z)^{\gamma -\alpha -\beta }F(\gamma -\alpha
,\gamma -\beta ,\gamma -\alpha -\beta +1,1-z)
\end{equation}
then the negative frequency mode for $q=-1$ reads
\begin{equation}
\label{28}\Theta _{(q=-1)-}=c_{-}\sin ^{ik_R}(z)\sin (\frac
z2)F(a,b,ik_R+\frac 12,(q+1)/2)
\end{equation}
whereas the positive energy mode for $q=+1$ takes the form
\begin{equation}
\label{29}\Theta _{(q=+1)+}=c_{+}\sin ^{ik_R}(z)\sin (\frac
z2)F(a,b,ik_R+\frac 12,(q+1)/2)
\end{equation}
where $c_{+}$ and $c_{-}$ are constants of normalization. Using the relation
(\ref{27}) and to the fact that
$$
\Theta _{(q=-1)-}=\frac{\Gamma (ik_R+\frac 12)\Gamma (-ik_R-\frac 12)}{%
\Gamma (\left( \frac \xi {\chi ^{1/2}}+k\right) )\Gamma (-\left( \frac \xi
{\chi ^{1/2}}+k\right) )}\frac{c-}{c_{+}}\Theta _{(q=+1)+}+
$$
\begin{equation}
\label{30}\frac{\Gamma (ik_R+\frac 12)\Gamma (-ik_R-\frac 12)}{\Gamma
(ik_R+\frac 12+\left( \frac \xi {\chi ^{1/2}}+k\right) )\Gamma (ik_R+\frac
12-\left( \frac \xi {\chi ^{1/2}}+k\right) )}\frac{c-}{d_{+}}\Theta
_{(q=+1)-}
\end{equation}
finally, with the help of the condition $\mid \alpha \mid ^2+\mid \beta \mid
^2=1$ for spin 1/2 particles, we obtain that the density of particles
created $\mid \beta \mid ^2$is given by the relation
\begin{equation}
\label{beta}\mid \beta \mid ^2=\frac{k_R^2+1/4}{\left( \frac \xi {\chi
^{1/2}}+k\right) ^2}\left| \frac{\Gamma (ik_R+\frac 12)\Gamma (-ik_R-\frac
12)}{\Gamma (\left( \frac \xi {\chi ^{1/2}}+k\right) )\Gamma (-\left( \frac
\xi {\chi ^{1/2}}+k\right) )}\right| ^2=\frac{\sin ^2\left( \frac \xi {\chi
^{1/2}}+k\right) \pi }{\cosh ^2\pi k_R}
\end{equation}
The expression (\ref{beta}) shows that the density of created particles has
a maximum for $\xi /\chi ^{1/2}=N+1/2,$ also we have that , there is not
particle creation when $\xi /\chi ^{1/2}$ takes integer values. This one is
a straightforward consequence of that the eigenvalues $k$ of the angular
momentum operator (\ref{K}) are also integers. Then, for $\xi =0$ we
reobtain the well known result that no particle creation is observed when
massless particles couples conformally to the metric.\cite{Zeldovich} The
particular form for $\mid \beta \mid ^2$is analogous to the one obtained in
analyzing quantum effects in a Robertson Walker metric with an expansion
factor $a(\eta )=\varepsilon \cosh ^{-1}\eta ,\cite{victor2,Grib1}$ as well
as the quantum mechanical problem of tunneling across a barrier of the form $%
U=-U_0$ $\cosh ^{-2}(\kappa x),$ where for given values of $\kappa $ the
reflection coefficient vanishes.

\vspace{.5cm}

\centerline{\bf Acknowledgments}

\noindent The author wishes to express his indebtedness to the Centre de
Physique Th\'eorique for the suitable conditions of work. Also the author
wishes to acknowledge to the CONICIT of Venezuela and the Vollmer
Foundation for financial support.


\newpage


\begin{thebibliography}{99}
\bibitem{parker}  L. Parker {\it Phys. Rev. Lett}. {\bf 21} 562 (1968)
\bibitem{Grib}  A. A. Grib, S. G. Mamaev, and V. M. Mostepanenko, Sov. J.
Nucl. Phys {\bf 10 }722 (1970)
\bibitem{Bertotti}  B. Bertotti, {\it Phys Rev}. {\bf 116} 1331 (1959)
\bibitem{Robinson}  I. Robinson, {\it Bull Acad. Pol. Sci}. {\bf 7} 351
(1959)
\bibitem{sahni}  V. Sahni , {\it Phys. Lett. A} {\bf 86}, 87 (1981)
\bibitem{Zeldovich}  Ya. B. Zeldovich and A. A. Starobinskii {\it Sov. Phys.
JETP} {\bf 61} 2161 (1971)
\bibitem{Shishkin1}  G. V. Shishkin and V. M. Villalba {\it J. Math. Phys. }%
{\bf 33} 2093 (1992)
\bibitem{Shishkin2}  G. V. Shishkin and V. M. Villalba {\it J. Math. Phys. }%
{\bf 34} 5037 (1993)
\bibitem{Brill}  D. Brill and J. A. Wheeler Rev. Mod. Phys. {\bf 29} 465
(1957)
\bibitem{victor2}  V. M. Villalba and U. Percoco, {\it Can. J. Phys} {\bf 70
}143 (1992)
\bibitem{Grib1}  A. A. Grib, S. G. Mamaev, and V. M. Mostepanenko. {\it %
Quantum vacuum effects in strong fields} (Energoatomizdat 1988)
\end{thebibliography}
\end{document}